\documentclass{llncs}
\usepackage[english]{babel}
\usepackage[version=0.96]{pgf}
\usepackage{tikz}
\usetikzlibrary{arrows,shapes,snakes,automata,backgrounds,petri}
\usepackage[latin1]{inputenc}
\usepackage{todonotes} % Todo messages in the margins
\usepackage{amsmath}
\usepackage{amsfonts}
\usepackage{shuffle} % \shuffle operator
\usepackage{float}
\usepackage{array}
\usepackage[compatibility=false]{caption}
\usepackage{graphicx,subcaption,mwe}
\usepackage{multirow}
\usepackage{bm}
\newcolumntype{P}[1]{>{\centering\arraybackslash}p{#1}}

\usepackage{verbatim}
\usepackage{hyperref}
\newcommand{\multiset}{M}
\title{Prototype Selection Based on Clustering and Conformance Metrics for Model Discovery}
\author{Mohammadreza Fani Sani\inst{1} \and Mathilde Boltenhagen\inst{2} \and Wil van der Aalst\inst{1,3}  }
\institute{RWTH Aachen University, Aachen, Germany\\
\email{\{fanisani, wvdaalst\}@pads.rwth-aachen.de}
\\
\and
 LSV, Universit\'e Paris-Saclay, ENS Paris-Saclay, CNRS, Inria, Cachan (France)
\\
\email{\{boltenhagen\}@lsv.fr}
\and
Fraunhofer FIT, Birlinghoven Castle, Sankt Augustin, Germany 
\\	
}

\begin{document}

\sloppy

\maketitle

\begin{abstract}
	Process discovery aims at automatically creating process models on the basis of event data captured during the execution of business processes.  
	Process discovery algorithms tend to use all of the event data to discover a process model.
	This attitude sometimes leads to discover imprecise and/or complex process models that may conceal important information of processes. 
	To address this problem, several techniques, from data filtering to model repair, have been elaborated in the literature.
	In this paper, we introduce a new incremental prototype selection algorithm based on clustering of process instances.
	The method aims to iteratively compute a unique process model with a different set of selected prototypes, i.e., representative of whole event data and stops when conformance metrics decrease. 
	The proposed method has been implemented in both the ProM and the RapidProM platforms.
	We applied the proposed method on several real event data with state-of-the-art, process discovery algorithms. Results show that using the proposed method leads to improve the general quality of discovered process models. 
	
\end{abstract}

\section{Introduction}
\textit{Process Mining} bridges the gap between traditional data analysis techniques, like data mining, and business process management analysis \cite{aalst_2016_pm}. 
 Process discovery, one of the main branches of the field, aims to discover process models (commonly Petri nets or BPMN) that accurately describe the underlying processes captured within the event data \cite{aalst_2016_pm}. 
 Process models capture choice, concurrent and loop behavior of activities.
 
To measure the quality of discovered process models, four criteria have been presented in the literature, i.e., \textit{fitness}, \textit{precision}, \textit{generalization} and \textit{simplicity}~\cite{buijs_2012_role}. 
Fitness, that seems to be the most well addressed criterion, indicates how much the observed behavior of the data is described by the process model.
In opposite, precision describes over language of the model, i.e., it computes how much modeled behaviors indeed exist in the event log. 
Generalization aims to quantify the flexibility of process model to describe behavior that is not presented in the event log but possible in the process.
Simplicity measures the understandability of a process model by limiting the number of nodes and complex structures of the resulted model. 

Many process discovery algorithms have been proposed in the literature. 
However, when dealing with the complexity of real data, they face problems to discover proper models.
Many algorithms tend to depict most or even all of the process instances and create perfectly fitting process models. 
For large event logs, resulting models are then too complex and often imprecise. Thus, the main problem of many state-of-the-art, process discovery algorithms is to balance between these four quality criteria. 
To deal with the quality metrics, some research incorporates conformance checking artefacts in process discovery algorithms which naturally bring better results. Genetic algorithm based approaches like \cite{de2007genetic,van2014genetic} have been proposed but they are time costly.

While many algorithms proposed recently, a novel approach to improve process models quality has emerged: \emph{data preprocessing}.  
Several filtering methods have been presented \cite{conforti_2017_filtering,Fani_Filtering_2017}. 
Moreover data clustering have showed good results in order to get several simpler models \cite{song2008trace,bose2009context,de_2013_active,tax2016mining,boltenhagen2019generalized}. 
The quality of the simpler process models to their assigned traces is then better than a unique large model.
 However, decision makers prefer a unique visualization of their system. 
 %It is clear that by increasing the number of clusters, we increase the quality of each process models, but the complexity of interpreting many process models is also higher. 

In this paper, we address the quality issues of process discovery algorithms and propose a general incremental prototype selection algorithm based on clustering and conformance artefacts. To get a unique process model, our method uses trace clustering in order to get one sublog of selected traces as a representative of the whole event log. This sublog, called prototypes, allows one to reduce the data variability and, by using the prototypes as input of the discovery algorithms, improves quality of the discovered model. The prototype selection is incremental and depends on a moderate use of conformance checking artefacts. 

Using the \texttt{ProM}-based \cite{verbeek_2010_xes} extension of \texttt{RapidMiner}, i.e., \texttt{RapidProM}~\cite{Rad}, we study the usefulness of the proposed method using real event logs in combination of different process discovery algorithms.
The experimental results show that applying our method increases the balance between the quality metrics of discovered process models.

The remainder of this paper is structured as follows. 
In \autoref{sec:related_work}, we discuss related work.
\autoref{sec:preliminaries} defines preliminary notation. 
{We present the proposed prototype selection algorithm in} \autoref{sec:Method}.
The evaluation and corresponding results are given in \autoref{sec:eval}.
Finally, \autoref{sec:conclusion} concludes the paper and presents some directions for future work.

\section{Related work}
\label{sec:related_work}
Quality of discovered process models depends on data which can be noisy or very complex \cite{bose2013wanna}. Process discovery algorithms face problems when dealing with very heterogenous process event log and generate spaghetti-like process models, i.e., the discovered models contain too many nodes and arcs. 
It occurs that discovered structures are too dense for human analysis.
 To reduce event logs to only significant behaviors, different variants of process discovery algorithms like the Inductive miner \cite{leemans2013discovering} have been developed to filter infrequent traces. 
 Those traces can be presented as outliers and should be filtered \cite{ghionna2008outlier}.
 There are research, e.g., \cite{Fani_SFiltering_2018,conforti_2017_filtering,Fani_Filtering_2017} that aim to increase the quality of process models by removing outlier behavior.
  In \cite{buijs_2012_role}, the authors proposed a genetic algorithm based method that aims to discover a process model with the highest possible quality metric that is defined by the user.
Furthermore, works like \cite{sani_2019_impact,bauer_2018_much} already presented sublog selection in order to improve model discovery. \cite{sani_2019_impact} is based on sequences frequency while \cite{bauer_2018_much} does sampling. 

In the other hand, several works \cite{song2008trace,bose2009context,tax2016mining,boltenhagen2019generalized,de_2013_active} focused on getting many sub-models on the same event log by using clustering method.
 Traces are clustered into sub-logs to obtain a reasonable number of differences between the process instances in the same cluster.
 Thus, the resulting sub-logs of the different clusters are used to discover the sub-models. 
 The quality of the sub-models are greater than a general process model discovered with the whole event log.
 Different clustering methods are used to get similarities between log traces in order to group them. 
 \cite{song2008trace} proposes an approach that combines the resources of data from occurrence of activities to meta-information in usable vectors. In \cite{bose2009context}, similarities between traces are computed as distances between the sequences of activities, more precisely edit distance. Closer to \emph{Pattern Sequential Mining}, authors of \cite{tax2016mining} use clustering to discover several models that describe existing local patterns. 
  This paper also highlights the readability of the sub-models compared to the complete one. Indeed simplicity is still an issue of large databases. In \cite{boltenhagen2019generalized}, authors introduced a clustering method in which centroids are submodels. Finally, work of \cite{de_2013_active}, which is certainly the closer to ours, create clusters with an incremental method that incorporates fitness metric in order to get good log-conform sub-models. All the listed papers on clustering return several process models which may be a barrier for decision makers who need a unique overview. 

Furthermore, our work reconsiders the definition of F-measure for Process Mining which has already been considered by \cite{de_2011_robust} that proposes a variant based on artificial generative negative events. 

\section{Preliminaries}
\label{sec:preliminaries}
\begin{figure}[tb]
	
	\centering
	\includegraphics[width=0.85\linewidth]{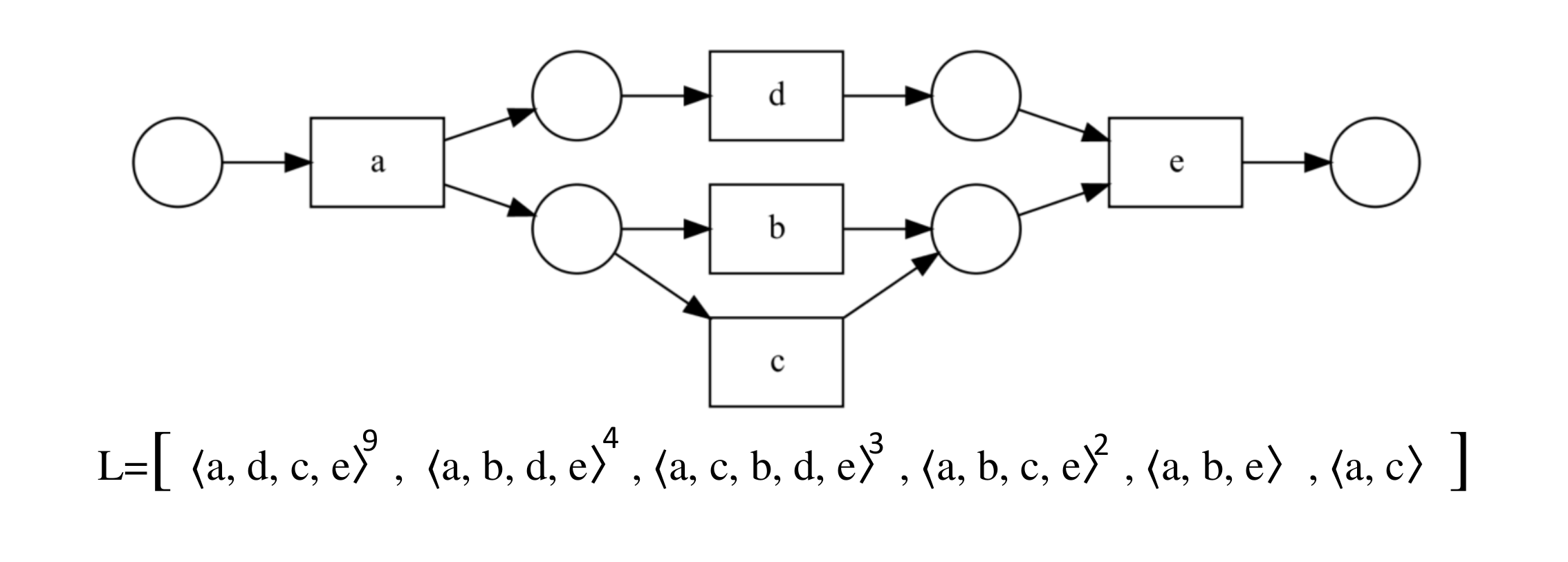}
	\caption{An example process model with Petri net notation and an event log. }
	\label{fig:Example}
\end{figure}

In this paper, we focus on sequences of activities, i.e., also called traces, structured like words. Then an event log is defined as follows. 
\begin{definition}[Event Log]
	Let $\mathcal{U}_A$ be a set of activities.
	An event log is a multiset of sequences over $\mathcal{U}_A$, i.e., $L \in \multiset(\mathcal{U}_A^*)$ that is a finite set of words.
 A word in an event log is also called \emph{log traces}.
 
\end{definition}
For instance, in \autoref{fig:Example}, an example event log $ L $ is presented with six unique log traces.
The occurrence frequency of the word (or trace) $ \langle a, d, c, e \rangle $ is nine. 

One can create $ l $ as a sublog of an entire event log $ L $ that $ l \subseteq L $. 
It is possible to separate a log to different soblogs using clustering methods.
\begin{definition}[Trace Clustering]
 Given a log $L$, a trace clustering $\xi(L,n)$ is a partitioning of L in a set of sublogs $\{ l_{1}, l_{2} \dots l_{n} \}$ such as $\forall_{i \not = j} \{l_{i} \cap  l_{j}\} = \emptyset $ and  $\displaystyle{ \biguplus_{i=1:n} l_{i} = L }$ . 
\end{definition}

There are different clustering methods. 
However, they commonly work based on a distance metric that returns how two items are different from each others. Many clustering algorithms like KMeans \cite{hartigan1979algorithm} create sublogs by considering average items called \emph{centroids}.

\begin{definition}[Centroids]
For a trace clustering $\xi(L,n)= \{ l_{1}, l_{2} \dots l_{n} \}$, \emph{centroids} is a set $\{c_1, \dots, c_n \}$ defined as $\forall_{i \in \{1\dots n\}} c_i = avg_{\xi}(l_{i})$ with $avg_{\xi} : \{ l_{1}, l_{2} \dots l_{n} \} \rightarrow \{c_1, \dots, c_n \} $ a cost function which relates sublogs to its centroid. Notice that $avg_{\xi}$ is usually defined by $\xi$.
\end{definition}

One distance metric that is widely used for clustering of words is the Levenshtein's distance also called edit distance. 

\begin{definition}[Edit distance]
	Suppose that $ \sigma,\sigma' \in A^* $, Edit Distance function $ \bigtriangleup (\sigma, \sigma')\to \mathbb{N}$ returns the minimum number of edits that are required to transform $ \sigma $ to $ \sigma' $.
\end{definition}
We assume that an edit operation can only be a deletion or an insertion of an activity (or a transition label) in a trace.
To give an example, $ \bigtriangleup(\langle a,c,f,e,d \rangle,$ $ \langle a,f,c,a,d \rangle )=4 $ corresponding to two deletions and two insertions.

We are able to define a process model $ M $ as the set of all words (i.e, traces) that it describes. 
For example, the set of all possible words of that can be presented by this model is $ \{\langle a, b, d, e   \rangle, \langle a, d, c, e   \rangle, \langle a, c, d, e   \rangle, \langle a, d, b, e   \rangle      \} $.
 
We define that one log trace $ \sigma_L $ in $ L $ is fitted in the process model $ M $, if it $ \sigma_L \in M$. 
If a trace is not fitted to a Model, we compute its similarity by the following formula.
\begin{equation}
fitness (\sigma_L, M)= \frac{\displaystyle{\min_{\forall \sigma \in M }} \bigtriangleup(\sigma_L, \sigma) }{|\sigma_L|+\displaystyle{\min_{\forall \sigma \in M} |\sigma|}}
\end{equation} 
The above equation returns a value between 0 and 1 that the value 1 refers to a completely fitted trace. 

\section{Incremental Prototype Selection Method for Process Discovery}
\label{sec:Method}

\begin{figure}[tb]
	\centering
	\includegraphics[scale=0.52]{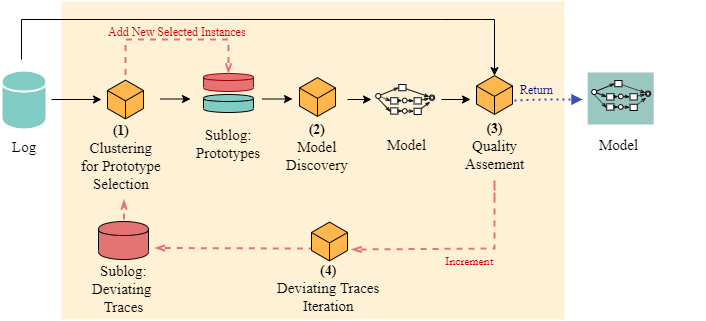}
	\caption{Structure of the Prototype Selection Approach}
	\label{method summary}
	\vspace{-0.2cm}
\end{figure}

In this section we bring the details of our approach.
The schematic view of the proposed method is given in Fig.~\ref{method summary}. 
This method contains the following four main steps:
\begin{enumerate}
	\item \emph{Clustering for prototype selection}: by using a clustering method, we select prototypes.
	\item \emph{Model discovery}: the method applies a process discovery algorithm on the selected prototypes. 
	\item \emph{Quality assessment}: to relate the original event log to the discovered model, conformance artefacts are computed.
	\item \emph{Iteration over deviating traces}: while quality metrics improve, we iterate the process on deviating traces of the last discovered model.
	
\end{enumerate}

The selection of prototypes is then an iterative process and the sublog of selected prototypes gently grows at each iteration that permits to increase fitness while the precision value decreasing. 
In the following we explain each of the above steps

\subsection{Prototype Selection}
\label{Initialization : Selection of Instances }
By applying process discovery algorithms directly on whole real event logs, we usually have complex and imprecise process models. 
As presented in \cite{sani_2018_repairing,sani_2019_impact} by modifying and sampling event logs we are able to improve results of process discovery algorithms in terms of F-measure.
A key contribution of our work is to select prototypes by using a clustering method in this regard. 
However, it is possible to use other policies to select prototypes of an event log.
\subsubsection{Clustering Approach}

As presented in \autoref{sec:preliminaries}, a trace clustering method separates traces in sublogs by considering their similarity. Similarity of two traces can be defined over activities occurrences, activities frequency, resources attributes or the order and sequence of activities. 
In this paper, the last criterion is chosen.
Therefore, we propose to cluster traces by comparing the distances between activities orders, more precisely the edit distance. 

Unlike previous works, we apply clustering to extract a small set of representatives traces. 
Some clustering algorithms like \emph{K-Medoids} \cite{de_2013_effective} (i.e., a variant of KMeans) determines the centroid of each cluster in the input, i.e., in traces clustering, centroids are traces containing in the event log. 
From this assumptions, we defined \emph{prototypes}. 

\begin{definition}[Prototypes]
	For an event log $L$, a trace clustering method $\xi$ and a number of clusters $n$, such as \(\xi(L,n) = \{ l_{1}, \dots , l_{n}\} \) and $\{c_1, \dots, c_n\}$
	 are the centroids, such as $\forall_{i \in \{0\dots n\}} c_i = avg_{\xi}(l_i)$.
	 $\{c_1, \dots, c_n\}$ are 
	 defined as \emph{prototypes} if $\{c_1, \dots, c_n\} \in L$. 
\end{definition}

That means a prototype is a trace as a centroid of a cluster.
The number of clusters and consequently the number of prototypes, is an input of the method that is defined by the user. 
The prototypes are then a small set of traces that represent the whole event log.

\subsection{Model discovery}
After finding the prototypes that are described above, we discover a descriptive view of it. 
In this regard, we are flexible to use any process discovery algorithm.
However, it is recommended to use methods that generate sound process models. 
By discovering a model from the selected prototypes, we will have a general view of what is going in the process and position different log traces w.r.t, this model. 

\subsection{Quality assessment}
\label{subsec:Conformance Checking Incorporation }
At this point, we have a discovered model from a small set of representative traces, i.e., prototypes. 
To ensure that the process model conforms to the whole event log, we incorporate quality assessment evaluations in our method. The metrics are then computed by considering the original event log and not just the prototypes. 

As explained before, there are different metrics to evaluate the quality of a process model.
The proposed method uses common fitness metric described in section \ref{sec:preliminaries} and the ETC precision that is presented in \cite{munoz_2010_fresh}. 
However, we are able to use other metrics too. 
The classical F-measure that is used in Process Mining domain, levels fitness and precision with equal weights as defined by the following equation:

\begin{equation}
\label{equ:FM}
F = \frac{2 * Precision * Fitness}{Precision+Fitness}
\end{equation}

However, from the user interests, fitness and precision may have different importance. To adapt weight on the desired result, we use a variant of the F-measure entitled the $\beta$ F-Measure introduced in \cite{chinchor1992muc} that is a more genera definition of \autoref{equ:FM}:
\begin{equation}
\label{equ:FMB}
F_{\beta} = (1 + \beta^2 ) * \frac{ Precision * Fitness}{(\beta^2 *Precision)+Fitness}
\end{equation}

This variant allows us to weight precision by setting the variable $\beta \in [0,+\infty]$. For $\beta > 1$, one raises fitness importance and for $\beta < 1$ precision is put forward. $\beta$ attribute is the third and last parameter of our method, i.e., the number of clusters and the chosen process discovery algorithm are the two first parameters. 
For instance, using $ F_1$-Measure means that $ \beta =1 $ and both the precision and fitness have the same weight and in $ F_2$-Measure the weight of fitness is higher.

\subsection{Incremental Method and Return Condition }

The $F_{\beta}$ -measure is computed for the first time in our method after the initialization step that selects a first set of prototypes. 
At this point, the proposed method starts an iterative procedure as follows.
At first, in each iteration, the method finds the deviating traces that are formally defined as follows.

\begin{definition}[Deviating Traces]
	From a process model $M$ and an event log $L$, the \emph{Deviating Traces} is a subset $L_d \subset L$ such as $\forall_{\sigma \in L_d} \ fitness(\sigma, M) < 1$.
\end{definition}

After finding the deviating traces, we look for representatives of them like what we did in section \ref{Initialization : Selection of Instances }.
Then the new set of prototypes will be added to the previous one (see Fig.~\ref{method summary}). 
Thereafter, we apply the process discovery algorithm to find a new process model and so on. 
Loop stops by comparing previous and current $F_{\beta}$-measure.
 While conformance of the discovered model is getting better, the method tries to add a new set of selected prototypes. 

The number of prototypes raises at each recursion which usually implies a precision decreasing but a potential fitness improvement.
 $F_{\beta}$-measure balances the metrics from the user point of view with the $\beta$ parameter. In this part of our method, we see that the discovery algorithm is the key of the $F_{\beta}$ value. 
 We make the hypotheses that discovery algorithms tend to approach perfect fitness and adding traces in the input raises the fitness of the whole log and decreases precision. This hypotheses is commonly true (as also assumed in \cite{augusto_2019_metaheuristic}). 
Therefore, the algorithms stops when there is no improvement in $ F_{\beta} $ of discovered process model of prototypes.
The method returns both the discovered process model and prototypes.

\section{Experiments}
\label{sec:eval}

In this section, we aim to indicate possibility of improving the quality of discovered process models using the proposed method. 
We first explain the implementation of this method.
Afterwards, we present the data that are used in the evaluation and the experimental settings. 
Finally, we show evaluation results and discuss about our findings. 

\subsection{Implementation}
\label{sec:imp}
To apply the proposed method, we implemented the \textit{Prototype Selection} plug-in in \texttt{ProM} framework\footnote{Prototype Selection plug-in:\scriptsize\url{svn.win.tue.nl/repos/prom/Packages/LogFiltering}}.
The plug-in takes an event log as input and outputs the discovered model.

As presented above, our method works for different settings. Three parameters must be selected by the user. 
First, it is possible to choose the number of clusters and consequently the number of prototypes to build the process model. 
To cluster traces, we implemented the K-Medoids~\cite{de_2013_effective} algorithm that the number of the clusters is driven from input.
Then, we have a representative trace per cluster, i.e., its centroid which is a prototype of our method.
Afterward, the user has to define the $\beta$ parameter to apply different weights on fitness and precision metrics as explained in~\ref{subsec:Conformance Checking Incorporation }.
Finally, the last parameter is the discovery algorithm that will be run by the method.

In addition, we ported the \textit{Prototype Selection} plug-in to \texttt{RapidProM} that allows to apply our proposed method on various event logs with different parameters.
\texttt{RapidProM} is an extension of \texttt{RapidMiner} that combines scientific work-flows with a range of (\texttt{ProM}-based) process mining algorithms.

\subsection{Experiment Settings}
\label{sec:setting}
The experiments have been conducted on eight real event logs \footnote{ \url{https://data.4tu.nl/repository/collection:event_logs_real}} of different fields. e.g., health-care to insurance.
Event logs have different characteristics which are given in Table~\ref{tab:logs}. 

\begin{table}[t]
	\centering
	\begin{tabular}{|l|l|l|l|l|}
		\hline
		\textbf{Event Log} & \textbf{Activities\#} & \textbf{Traces\#} & \textbf{Variants\#} & \textbf{DF Relations\#} \\ \hline
		
		\textit{$ \text{BPIC}_-2012$}\cite{BPI_2012_event} & 23 & 13087 & 4336 & 138 \\ \hline
		
		\textit{$ \text{BPIC}_-2018_-\text{Dept.} $}\cite{BPI_2018_event} & 7 & 43808 & 59 & 12 \\ \hline
		\textit{$ \text{BPIC}_-2018_-\text{Insp.} $}\cite{BPI_2018_event} &15 & 5485 & 3190 & 67 \\ \hline
		\textit{$ \text{BPIC}_-2018_-\text{Ref.} $}\cite{BPI_2018_event} & 6 & 43802 & 515 & 15 \\ \hline
		\textit{$\text{BPIC}_-2019$}\cite{BPI_2019_event} & 44 & 251734 & 11973 & 538 \\ \hline
		\textit{$ Hospital $}\cite{mannhardt_2017_hospital} & 18 & 100000 & 1020 & 143 \\ \hline
		\textit{$ Road $} \cite{de_2015_road}& 11 & 150370 & 231 & 70 \\ \hline
		\textit{$ Sepsis $}\cite{Sepcis_2016_Felix} & 16 & 1050 & 846 & 115 \\ \hline		
	\end{tabular}
		\caption{Details of real event logs that are used in the experiment}
		\label{tab:logs}
\end{table}

As the \emph{Prototype Selection} has many parameters, i.e., the number of clusters, the $\beta$ value and the discovery algorithm, we show results over a set of different settings.
We repeated the experiments for $ 2 $ to $ 9 $ clusters and set different $ \beta $ values in $ \{ 0.5, 1 , 2\} $ to compute F-Measure. 

For process discovery, we used the Inductive Miner~\cite{leemans_2014_ind_infreq}, the ILP miner~\cite{vanZelst2017_comp}, and the Split Miner~\cite{augusto_2019_split}.
As the Inductive Miner and the Split Miner have internal settings too, we first compare in section~\ref{sec:exp Qualitative} our method by using only the ILP Miner and show a complete overview of the resulted quality improvement.
Then, in section~\ref{sec:exp quantitative} we complete the experiments with 50 different settings for the Inductive Miner (IMi) and 100 for the Split Miner. 

Moreover, we compared our works to related works, i.e., \textit{Sampling}~\cite{sani_2019_impact} and \textit{Statistical}~\cite{bauer_2018_much} methods, that, as the proposed method, also selects some of process instances in logs.
For both of these methods, we used the same settings that are explained in \cite{sani_2019_impact}.
We also compared our method to normal process discovery algorithms, i.e, discovery without preprocessing denoted \textit{Nothing} in the experiments.
However, notice that the internal filtering mechanisms of process discovery algorithms are still used.

\subsection{Qualitative Experiments}
\label{sec:exp Qualitative}

As the aim of our method is to highlight improvement in most of the process mining criteria, we first present Fig.~\ref{tab:Average} that figures different measures like fitness, precision and simplicity by using the ILP miner.
For simplicity, we consider two metrics that measure the complexity of discovered process models. 
\emph{Size} of process models is a combination of number of transitions, places and arcs that connected them. 
Another metric is the Cardoso metric~\cite{lassen_2009_complexity} that measures the complexity of a process model by its complex structures, i.e., \textit{Xor}, \textit{Or}, and \textit{And} components. 
For both of these measures a lower value means less complexity and consequently a simpler process model.
For each event log, the best $ F_1$-Measure value is presented in bold.

\begin{table}[t]
	\centering
	
	\begin{tabular}{c}
		\includegraphics[width=\linewidth]{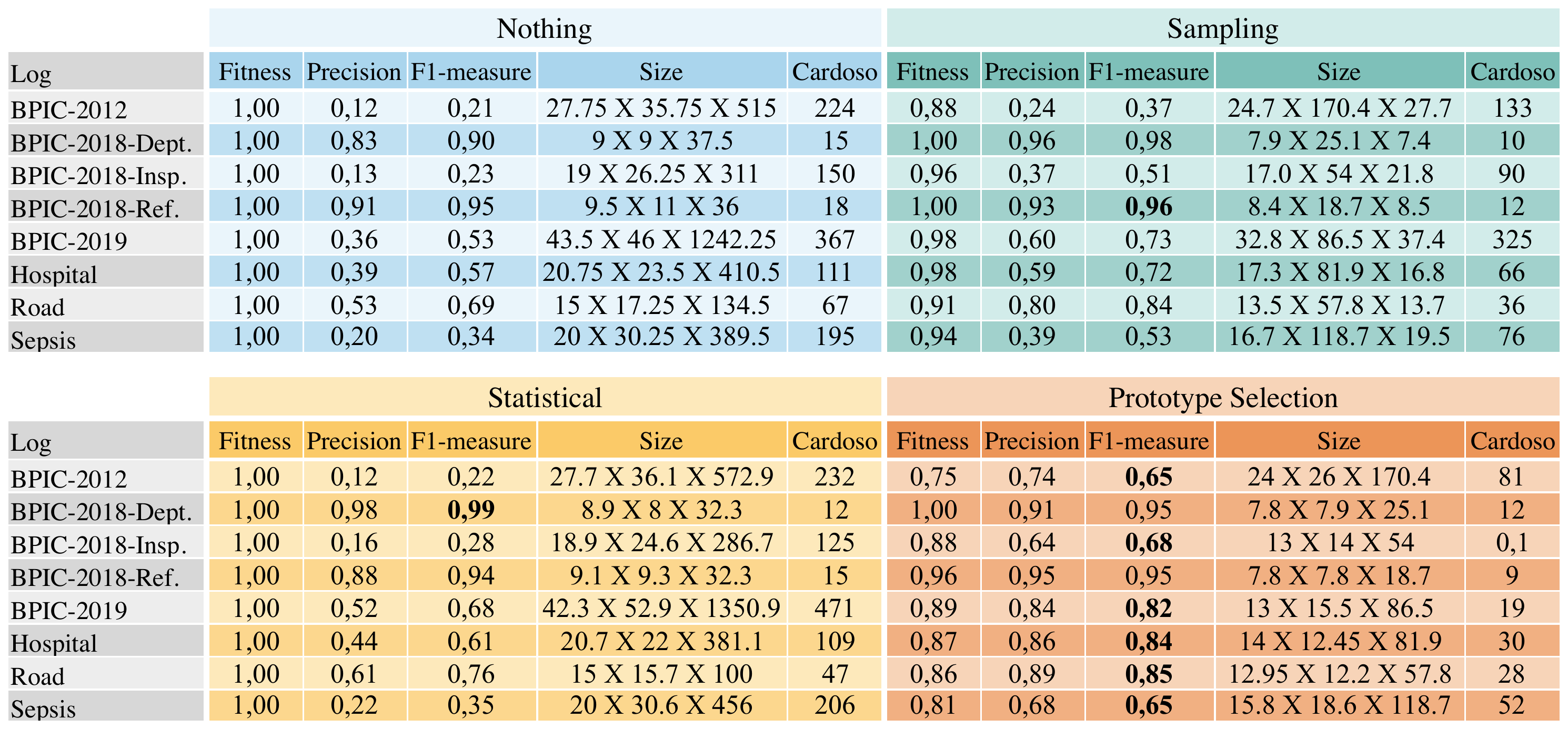}
		\end{tabular}
	\caption{Average values of quality criteria measures per preprocessing method for different event logs and process models discovered by using the ILP miner}
	\label{tab:Average}
	\vspace{-0.6cm}
\end{table}	

According to Fig.~\ref{tab:Average}, the \textit{Prototype Selection} method results in process models with higher $ F_1 $-Measure value and at the same time with lower complexity measures. 
The \textit{Statistical} method usually returns a process model close to the case that we do not use any preprocessing method. 
As process discovery algorithms usually result in high fitness, those cases have high fitness values. 
It happens because \textit{Statistical} tries to sample as much as possible unique behavior (i.e., trace variants). 
In other words, the goal if this {Statistical} method is to have smaller sublog similar to the original event log.
As expected, \textit{Sampling} and \textit{Prototype Selection} methods have lower fitness values.
Generally, those methods show a decrease in fitness value and increase in precision value which is the aim of the methods. 
However we can see that the \textit{Prototype Selection} brings a better balance on these values for large and complex event logs. 
As this table gives average of measures, fitness seems to be worst for \textit{Prototype Selection}. 
However, in this section, we focus on $\beta=1$ that gives the same weight to \textit{fitness} and precision.

For simple logs, e.g., $\textit{BPIC}-2018-\textit{Department}$~\cite{BPI_2018_event} that contains only $ 12 $ unique trace variants, the \textit{Sampling} method return better process models in term of $F1-$Measure. 
However, even for these event logs the quality of discovered process models using our proposed method is high. 
Furthermore our approach results in simpler process models (w.r.t, both complexity metrics) for all event logs that leads to have the higher understandability of undergoing process.

\begin{figure}[tb]
	\centering
	\begin{subfigure}[tbb]{\textwidth}	
		\includegraphics[width=\linewidth]{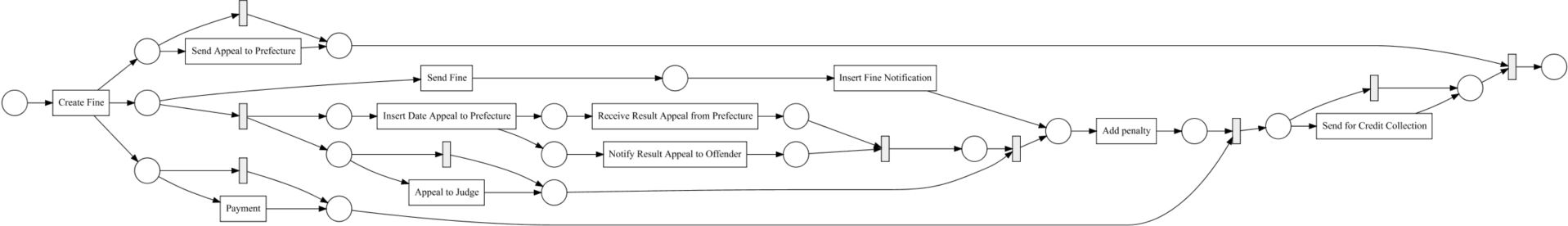}
		\caption{A process model that is discovered using the Inductive miner with the threshold value equals to $ 0.4 $ on the whole event log.  }
		\label{fig:RoadIMi}	
	\end{subfigure}
	\begin{subfigure}[tbb]{\textwidth}	
		\includegraphics[width=\linewidth]{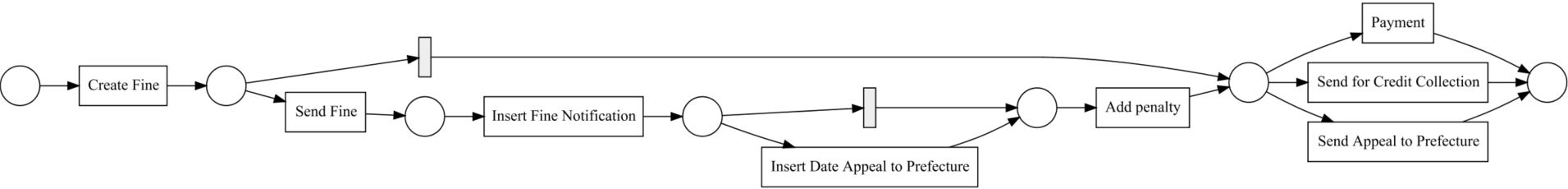}
		\caption{A process model discovered by the basic Inductive miner on the selected prototypes.  }
		\label{fig:RoadProto}
	\end{subfigure}

	\caption{ Comparing the process models that are discovered using the embedded filtering mechanism in the inductive miner and the \textit{Prototype Selection} method for the Road event log.  }
	\label{fig:Road}
\end{figure}

To give a better understanding of the outcome of the proposed method, we compare a process model that is discovered with the \textit{Prototype Selection} and the embedded filtering mechanism of the Inductive miner algorithm for the Road event log~\cite{de_2015_road} in \autoref{fig:Road}.
The fitness value of \autoref{fig:RoadIMi} is 0.78 and its precision is 0.65, however, these values are 0.92 and 0.94 for \autoref{fig:RoadProto}. 
Also, using the proposed method the discovered model is simpler. 
Note that three of infrequent activities are removed in \autoref{fig:RoadProto} as traces with those activities are not selected in prototypes. 
Therefore, by using our proposed method we are able to show a more general view of process as the selected prototypes are those that are most similar to the whole traces in the event log.

\subsection{Quantitative Experiments}
\label{sec:exp quantitative}
 
\begin{figure}[t]
	\centering
	\includegraphics[height=7cm,width=\linewidth]{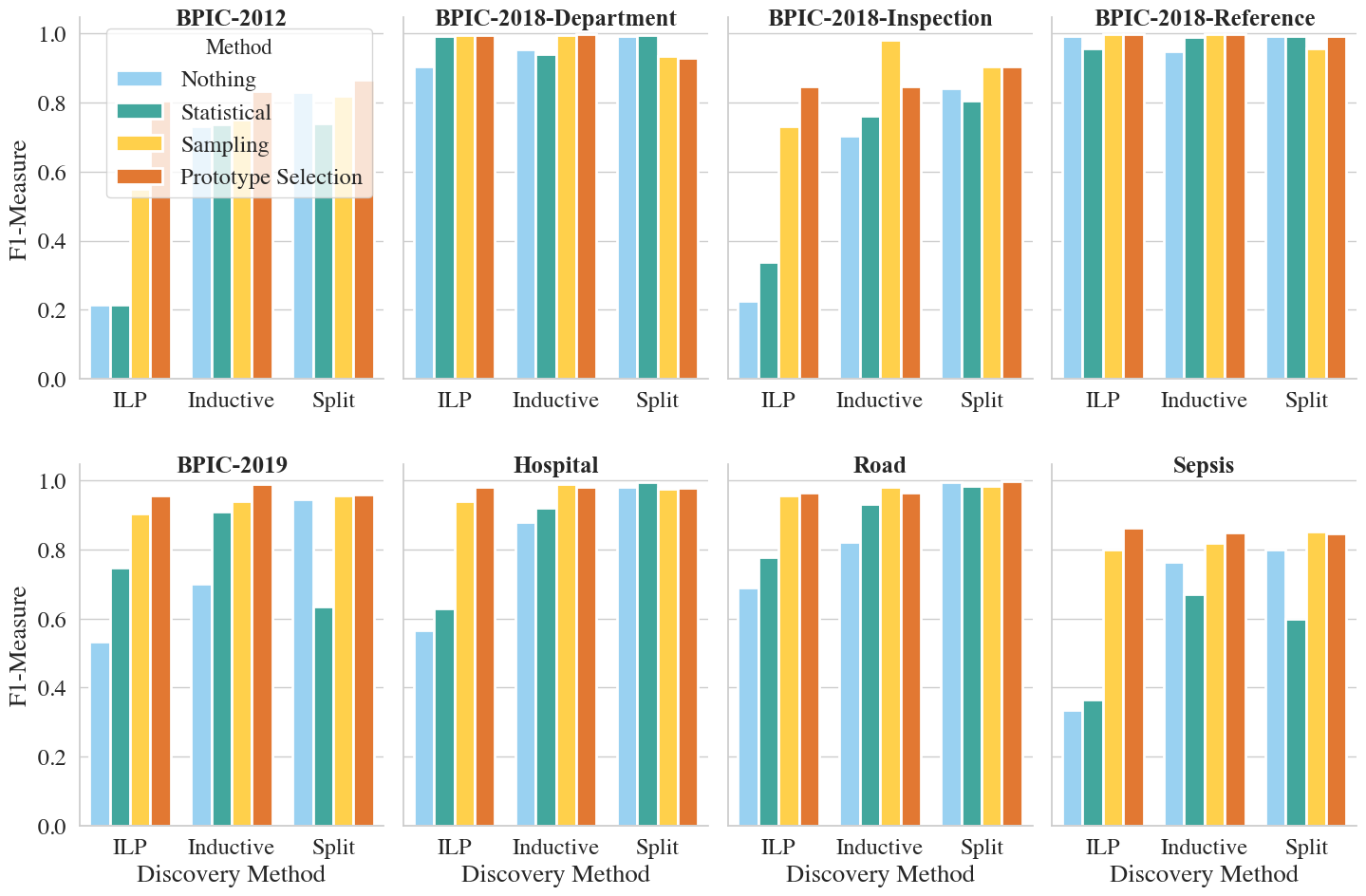}
	\caption{The maximum $F_1$-Measure value of different methods.  }
	\label{fig:F1}
\end{figure}

In this section, we computed $F_{\beta}-$measures of models discovered over three different algorithms and, for each of them, a set of settings (respectively 50 and 100 different settings for the Inductive Miner and the Split Miner). With Fig.~\ref{fig:F1} we show the \emph{best} $F_1$-measure of each algorithm and preprocessing method. 
Indeed, like the proposed method, \textit{Statistical} and \textit{Sampling} methods have different parameters that causes different outputs. 

It is shown that the proposed method, i.e., the Prototype Selection, in most of the cases results in the best solution or very close to it.
Specifically, when we use the ILP miner, the proposed method increases the $F_\beta$-Measure value more than statistical and sampling methods.
However, this improvement is not very impressive for the Split miner.
Note that the proposed is able to decrease the complexity of discovered process models for the Split miner.

\begin{figure}[t]
	\centering
	\includegraphics[width=\linewidth]{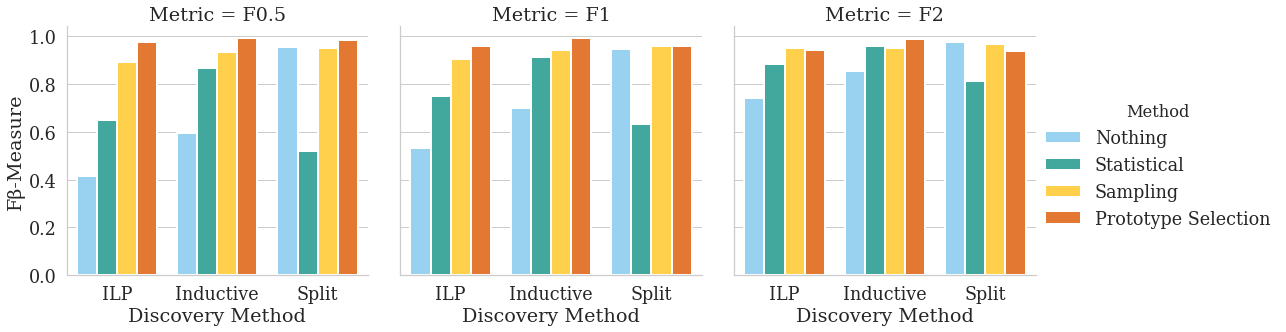}
	\caption{$F_{\beta}$-measure comparison of $\textit{BPIC}-2019$. }
	\label{fig:beta}
	\vspace{-0.4cm}
\end{figure}

In Fig.~\ref{fig:beta}, we show how the beta value influences over the results.
The charts figures three settings of event log $\textit{BPIC}-2019$~\cite{BPI_2019_event}.
 When one wants to work on more precise models, $\beta<1$ gives higher importance to precision metric. 
 In this case, we see that our method, which incorporates the $\beta$ parameter, returns much better results than the other preprocessing methods. 
 When fitness is preferred, i.e., $\beta>1$, we see that we cannot always determine the best method. 
 This is due to algorithm setting. 
 With this comparison, we want to highlight use of our method, many organizations need human understandable models and metrics balance for fitness and precision.

To conclude this subsection, we want to notify that the proposed method is discovery algorithm independent, i.e., we presented good results for different discovery algorithms. 

\subsection{Discussion}
Here, we aim to discuss how by selecting prototypes using the clustering method we improve the quality of process models.
We increased the number of prototypes and analyze the quality of the corresponding process model.
In this regard, Sepsis~\cite{Sepcis_2016_Felix} that contains lots of unique trace-variants and Road~\cite{de_2015_road} which have some dominant frequent variants were used. 
To discover process models, we used the basic Inductive miner. 
For this experiment, we used the cluster size equals to 1 that is different with the previous experiment.
We compared the case that we used centroids of clusters as prototypes or selecting most frequent variants.

\begin{figure}[t]
	\centering
	\includegraphics[width=\linewidth]{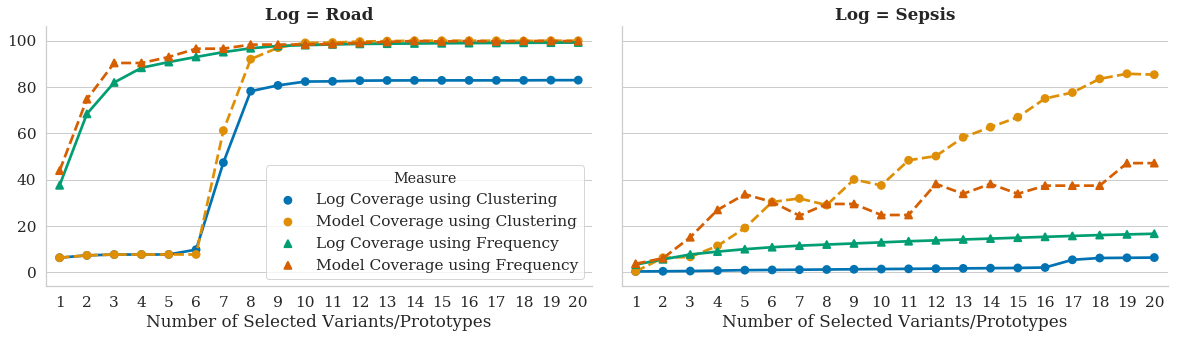}
	\caption{Effects of increasing the number of selected prototypes on the coverage of the discovered process models using frequency and clustering methods. }
	\label{fig:Investigate1}
	\includegraphics[width=\linewidth]{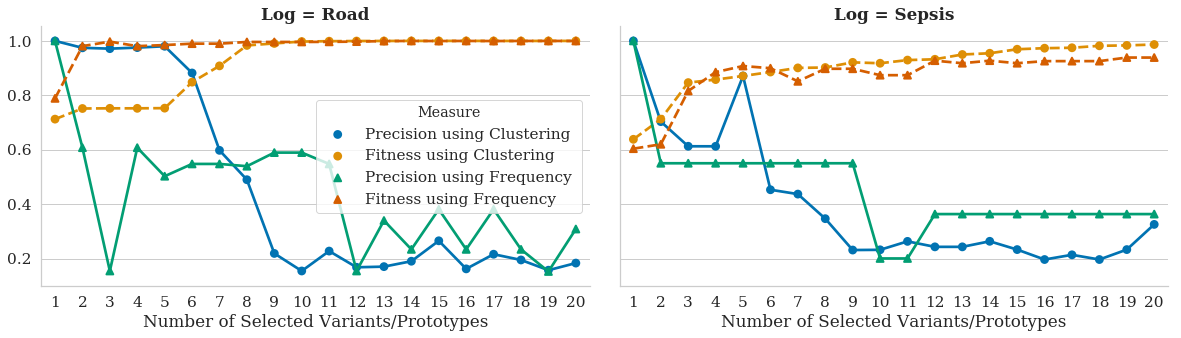}
	\caption{Effects of increasing the number of selected prototypes on the quality issues of discovered process models frequency and clustering methods. }
	\label{fig:Investigate2}
	\vspace{-0.4cm}
\end{figure}

Results of this analysis is shown in \autoref{fig:Investigate1}.
In this figure, the log coverage shows how many percentage of the traces in the event log, is corresponds to the selected prototypes/variants.
It is clear that the highest log coverage is achieved by selecting most frequent variants. 
Moreover, the model trace coverages indicates that how many percentages of traces in the event log is replayable (or perfectly fitted) by the discovered process model. 
For example, in the Sepsis event log, by selecting eight prototypes, i.e., corresponds to $ 5\% $ of traces, we are able to discover a process model that is able to perfectly replay 35\% of the traces in the event log. 
\autoref{fig:Investigate1} shows that process discovery algorithms depict much behavior in the process model compared to the given event log.
For event log with high frequent traces e.g., \textit{Road}, when we select few high frequent variants, we usually have higher model coverage. 
However, for event logs with lots of unique variants, e.g., \textit{Sepsis} or when we select more than 10 prototypes, the model coverage of clustering method is higher. 

In \autoref{fig:Investigate2}, we see how by increasing the number of prototypes the fitness value of discovered process models will be increased.
However, we usually have low precision by increasing prototypes.
This reduction is higher, when we select based on frequency. 
This experiment also shows that we are able to discover a high fitted process model without giving just few prototypes to process discovery algorithms.
We did not show it here, but we saw that by increasing the number of prototypes we will also have more complex process models.

\section{Conclusion}
\label{sec:conclusion}
In this paper, we proposed an incremental method to select prototypes of the event logs in order to generate high quality process models.
It clusters the traces in the event log based on their control-flow distances. 
Afterwards, it returns the most representative instance for each cluster, i.e., the centroids. 
We discover a process model of those prototypes which is analyzed with common conformance metrics.
Then starts the recursion over the deviating traces.
 A novel set of prototypes is added in the process model discovery which improves fitness while decreasing precision. 
 To evaluate the quality of process models, we recommend the use of $F_{\beta}-$Measure that allows one to weight fitness and precision. 

To evaluate the proposed method, we have developed a plug-in in the \textsc{ProM} platform
and also ported to \textsc{RapidProM} and 
 have applied the proposed prototype selection method on eight real event logs. We compared it with other state-of-the-art sampling methods using different process discovery algorithms. 
The results indicate that the proposed method is able to select process instances properly and help process discovery algorithms to return process models with better balance between quality measures. Discovered models are less complex and, consequently, it improves the understandability of them. 
Another advantage of our method is that it is more stable in chosen settings of parameters and it returns process models with higher quality in average. 

Currently, we use the prototypes for discovery purposes.
As future work, we aim to use them for other applications, e.g., conformance checking and performance analysis. 
One limitation of our method is it may find a local optimum rather than the global optimum. 
We plan to recommend a solution to have adjustable number of cluster for both initiating phase and incremental steps.   

\vspace{-0.3cm}
\section*{Acknowledgement}
We thank Josep Carmona, Thomas Chatain and Sebastiaan J. van Zelst for comments and supports that greatly improved the work.
\vspace{-0.3cm}
\bibliographystyle{splncs}
\bibliography{bibliography}
\end{document}